\documentclass[english,nofootinbib]{revtex4}
\usepackage{mathptmx}

\usepackage[T1]{fontenc}
\usepackage[latin9]{inputenc}
\usepackage[letterpaper]{geometry}
\usepackage[pdftex]{graphicx}
\usepackage{amssymb}

\makeatletter

\usepackage{babel}

\makeatother

\usepackage{babel}

\begin{document}

\title{Isocurvature Perturbations and Non-Gaussianity of Gravitationally Produced Nonthermal Dark
  Matter}

\author{Daniel J.~H.~Chung}
\email{danielchung@wisc.edu}
\author{Hojin Yoo}
\email{hyoo6@wisc.edu}
\affiliation{{\small Department of Physics, University of Wisconsin-Madison} \\
 {\small 1150 University Avenue, Madison, WI 53706, USA}}
\begin{abstract}
Gravitational particle production naturally occurs during the
transition from the inflationary phase to the non-inflationary phase.
If the particles are stable and very weakly interacting, they are
natural nonthermal dark matter candidates.  We show that such
nonthermal dark matter particles can produce local non-Gaussianities
large enough to be observed by ongoing and near future experiments
without being in conflict with the existing isocurvature bounds.  Of
particular interest is the fact that these particles can be observable
through local non-Gaussianities even when they form a very small
fraction of the total dark matter content.
\end{abstract}
\maketitle

\section{Introduction}

Standard slow-roll inflationary models with a single dynamical field
degree of freedom (e.g. see the review article \cite{Lyth:1998xn})
cannot generate large local non-Gaussianities (NGs)
\cite{Maldacena:2002vr,Seery:2005wm,Acquaviva:2002ud,Creminelli:2003iq},
which have been widely discussed and speculated upon in the
context of the cosmic microwave background (CMB) data
(e.g.~\cite{Yadav:2007yy,Komatsu:2008hk,Fergusson:2008ra,Smith:2009jr,Yadav:2010fz,Komatsu:2010hc,Curto:2011zt})
and large scale structure data
(e.g.~\cite{LoVerde:2007ri,Slosar:2008hx,Afshordi:2008ru,Matarrese:2008nc,Sefusatti:2007ih,Dalal:2007cu,Shandera:2010ei,Smith:2010gx,D'Amico:2010ta,Tseliakhovich:2010kf}). Many
multifield mechanisms have been proposed to generate observably large
local NGs (e.g.~
\cite{Linde:1996gt,Lyth:2002my,Zaldarriaga:2003my,Huang:2008zj,Kumar:2009ge,Byrnes:2010em,Barnaby:2010sq,Gong:2011cd,Demozzi:2010aj,Peterson:2010mv,Suyama:2010uj,Langlois:2011zz,Byrnes:2010xd,Burgess:2010bz,Mazumdar:2010sa,Chen:2009zp,Alabidi:2010ba,Takahashi:2009dr,Huang:2008bg,Kawasaki:2008sn,Chambers:2008gu,Yokoyama:2007uu,Sasaki:2006kq,Barnaby:2006cq,Cogollo:2008bi}). Most
of these models utilize coherent condensate field degrees of freedom
instead of incoherent many-particle states.

In this paper, we explore the possibility that nonthermal dark matter
(DM) particles gravitationally produced during the phase transition
out of the quasi-de-Sitter phase of inflation
\cite{Chung:1998zb,Kuzmin:1998kk} generate observably large
NGs.\footnote{These particles are sometimes called gravitationally
  produced WIMPZILLAs.} These dark matter particles can be viewed as
the remnants of de Sitter (dS) temperature driven radiation during
inflation, and no non-standard ingredients are needed for the
inflationary scenario for the purposes of this work. The only
nontrivial model requirement is that the dark matter either be very
heavy and/or superweakly interacting.

This class of scenarios effectively possesses only three important
independent dimensionful parameters: the Hubble expansion rate during
inflation, the reheating temperature, and the dark matter mass. Hence,
the physics is dominantly controlled by only two of these parameters
since the third converts the other two into dimensionless numbers. We
choose these to be the dark matter mass $m_{X}$ and the reheating
temperature $T_{\rm RH}$.  The existing cosmological data constraining the
isocurvature perturbation amplitude and the dark matter abundance
place bounds on the allowed parametric range for these parameters. We
find that to generate large observable local non-Gaussianities
characterized by an effective $f_{NL}$ parameter of around 30, there
is an upper bound of $m_{X}\lesssim4H_{e}$, where $H_{e}$ is the
expansion rate at the end of inflation. We also find that $f_{NL}$
will be suppressed if $T_{\rm RH} \gtrsim 10^{6}$ GeV if the dark matter
is absolutely stable.%
\footnote{This mass scale has part of its origins from the maximum dark matter
abundance today.%
} Somewhat surprisingly, even when the $X$ particles make up a small fraction
of the total dark matter content while thermal relics make up the
rest of the dark matter, observably large non-Gaussianities may be
imprinted.

The isocurvature perturbations in this class of scenarios have been
studied previously \cite{Chung:2004nh}. We note that this was also
briefly considered in \cite{Linde:1996gt}, which arrived at a
pessimistic conclusion.  However, that paper did not consider the
model as carefully as \cite{Chung:2004nh}, which reached a more
realistic conclusion regarding the viability of such scenarios. The
purpose of this paper is to point out that within this framework,
large local non-Gaussianities can be generated with a single
$O(10^{-1})$ tuning of the dark matter mass.

The order of presentation is as follows.  In Sec.~II, we discuss the
class of dark matter and inflation models for which the current
non-Gaussianity computation is relevant.  In Sec.~\ref{sec:twopoint},
we present a computation of the two-point function, including the
cross-correlation function between the isocurvature and the curvature
components.  The computation of the bispectrum and a presentation of
detailed arguments as to how the local non-Gaussianity can be large is
given in Sec.~\ref{sec:bispectrum}.  In
Sec.~\ref{sec:numericalresults}, we check our general analytic
arguments by computing in detail numerically the observables in the
context of $m^2 \phi^2/2$ inflationary model.  We then close the main
body of the paper with a summary and conclusions in
Sec.~\ref{sec:conclusions}.  In Appendix \ref{sec:modefunc}, we
present an analytic approximation to the mode function during
inflation that accounts for the small deviation from the pure dS
phase.  Finally,
in Appendix \ref{sec:justificationofthebackground}, we justify how the
effective classical background variable about which the classical
perturbations are defined is given by the expectation value of the
quantum operator.
Throughout this work, our metric convention is $(+,-,-,-)$.

\section{Class of dark matter and inflationary models}

We begin by defining the class of dark matter and inflationary models
considered in this work.  One requirement is that the dark matter
field $X$ be sufficiently long lived to be a viable dark matter
candidate.  Since we are considering an isocurvature dark matter
scenario, another requirement is that $X$ is very weakly interacting
with thermalized Standard Model (SM) particles that are assumed to
arise from the inflaton decay chain.  Although it is straightforward
to include dark matter interactions that allow transformations to
different particles, we will assume that the self-annihilation (or
coannihilation) interactions of the dark matter are too weak to change
the dark matter number density appreciably after it is produced at the
end of inflation.  Next, we note that even if $X$ is sufficiently
weakly interacting as not to thermalize, it may be a byproduct of a
slow-roll inflaton decay with a strength larger than gravitational
strength.  In such situations, the isocurvature nature of the $X$
particles produced gravitationally will be made impure by the inflaton
decay contribution. To keep the analysis simple for the purposes of
this paper, we will assume that the decay contribution is
negligible. Finally, we will assume that $X$ is a boson minimally
coupled to gravity. We will explore the complexities that arise when
some of these requirements are relaxed in future work.

The simplest model that satisfies the above criteria has two scalar
fields minimally coupled to gravity as follows: \begin{equation}
  S=\int
  d^{4}x\sqrt{g}\frac{1}{2}\left[(\partial\phi)^{2}-2U(\phi)+(\partial
    X)^{2}-m_{X}^{2}X^{2}\right]+S_{\rm SM}[g_{\mu\nu},\{\psi_{\rm
      SM}\}]+S_{\rm RH}[g_{\mu\nu},\phi,\{\psi_{\rm
      SM}\}],\end{equation} where $S_{\rm SM}$ is the SM sector,
$S_{\rm RH}$ is responsible for reheating, and $\phi$ is the inflaton
realizing a slow-roll inflationary scenario with
$U(\phi)=m^{2}\phi^{2}/2$. We note that since particle production during
the dS to non-dS phase leads to a dS horizon temperature population of
SM one-particle states, one might naively expect a minimum reheating
temperature of $T_{\rm RH}\gtrsim H_{e}/(2\pi)$, where $H_{e}$ is the
Hubble expansion rate the end of inflation.  However, this is
incorrect since during the coherent oscillations period between the end of
the inflation and the radiation dominated epoch, the SM radiation dilutes
with respect to the inflaton energy density.

Since we will carry out numerical analysis of the mode equations, this
simple model is useful.  Furthermore, it is clear that the results
generalize to a large class of models where the interactions are very
small.  We note also that the requirement of the interactions being weak
enough to avoid thermalization does not require a particularly
stringent limit on the interaction couplings.  For example, for
typical $O(1)$ coupling strengths for self-annihilations, it is
well known that large values of $m_{X}$ will naturally lead to
the nonthermal DM behavior desired in this paper if
\begin{equation}
\left(\frac{200\mbox{TeV}}{m_{X}}\right)^{2}\left(\frac{T_{\rm max}}{m_{X}}\right)\lesssim1,
\end{equation}
where $T_{\rm max}\gtrsim T_{\rm RH}$ is the maximum effective
temperature reached during reheating \cite{Chung:1998ua}.  Of course,
one possible model-building obstacle with large masses is that in
situations with accidental global symmetries protecting the stability
of the particles, higher-dimension operators must be suppressed to
avoid early decay.  Nonetheless, many viable beyond SM (BSM) models that
contain superheavy DM candidates have been proposed
\cite{Ellis:1990iu,Benakli:1998ut,Hamaguchi:1998nj,Hamaguchi:1999cv,Coriano:2001mg,Kusenko:1997si,Han:1998pa,Dvali:1999tq,Cheng:2002iz,Shiu:2003ta,Berezinsky:2008bg,Kephart:2001ix,Kephart:2006zd}.

Given that the $X$ particles have negligible non-gravitational
interactions and minimal couplings to gravity, they can only be
produced gravitationally or through initial conditions. We consider
the dynamics of an inflationary patch whose Bunch-Davies vacuum
\cite{Bunch:1978yq} satisfies
\begin{equation}
\lim_{k\rightarrow\infty}\hat{\alpha}_{k}|BD,0\rangle=0,
\end{equation}
where $\hat{\alpha}_k$ is the annihilation operator associated with
the curvature perturbations which is approximately dominated by the
inflaton.
This vacuum is also assumed to satisfy the ``no-particle'' condition
of the adiabatic vacuum during inflation
\cite{Birrell:1982ix,Parker1969,Chung:1998zb,Chung:2003wn}:
\begin{equation}
\hat{a}_{k}|BD,0\rangle=0,
\end{equation}
where $\hat{a}_{k}$ is the annihilation operator associated with the
$X$ field. The stress-energy tensor is renormalized such that
\begin{equation}
\langle BD,0|\hat{T}_{\mu\nu}^{(X,\mbox{ren})}|BD,0\rangle=0,
\end{equation}
which in practice is accomplished by the normal ordering of the
creation-annihilation operators in the adiabatic vacuum basis. This
means that the classical initial condition dependent DM density vanishes.

Nonetheless, because the transition from quasi-dS phase of inflation
to the non-dS phase after inflation represents a non-adiabatic
transition, it is well known that non-negligible particle production
occurs through Bogoliubov mixing of the creation-annihilation
operators, giving rise to a significant DM abundance today
\cite{Chung:1998zb, Chung:1998bt}.  The physics of this particle production
mechanism is similar to that of Hawking radiation.  In the intermediate mass
case where $m_{X} \sim H_{e}$ with minimal gravitational couplings, we find numerically that
the $X$ energy density at the end of inflation can be
approximated as $\rho_x(t_e)\approx 10^{-2} H_e^4$, which leads to the
relic abundance of $X$ particles today to be
\begin{equation}
  \Omega_{X}h^{2}\approx 10^{-1}\left(\frac{H_{e}}{10^{12}\mbox{
      GeV}}\right)^{2}\left(\frac{T_{\rm RH}}{10^{6}\mbox{
      GeV}}\right).\label{eq:gravproddarkmatter}
\end{equation}
What is interesting
about this scenario is that although the classical picture of particle
production occurs at the end of inflation, the correlations that are
relevant at the CMB scale are set long before the bulk of the particle
production occurs. This is intuitively self-consistent from the
Heisenberg time-energy uncertainty considerations. Although
Eq.~(\ref{eq:gravproddarkmatter}) yields the simplest possible
scenario, we later generalize the situation to the case of mixed dark
matter contributions in which the total cold dark matter (CDM)
abundance is given by \begin{equation} \Omega_{\rm CDM}=\Omega_{\rm
    therm}+\Omega_{X},\label{eq:mixed}\end{equation} where $\Omega_{\rm
  therm}$ are thermal relics that have only adiabatic perturbations
and $\Omega_{X}$ are relics that have dominantly isocurvature
perturbations. In this case, we define\begin{equation}
\omega_{X}\equiv\frac{\Omega_{X}}{\Omega_{\rm CDM}},\end{equation} and
will scale some of our computations to generalize our results to a
wider class of scenarios.

We now comment further on the inflationary model relevant for the
above DM scenario. The main features of the inflationary model that
are numerically important for the isocurvature and non-Gaussianity
analyses are $H_{*}$ (the Hubble expansion rate when the modes of
interest leave the horizon), $H_{e}$, and $T_{\rm RH}$ (the reheating
temperature).  As we will see, the primary role of $H_{*}$ is to
determine the spectral index of the isocurvature spectrum,
$H_{e}<H_{*}$ controls the particle production, and $T_{\rm RH}$
partially controls the map between the comoving wave vector and the
physical momentum. We assume that there are curvature perturbations
from the inflaton sector with the right magnitude to approximately
explain the CMB spectrum. As we will see and as is well known, the
current observational limits require that the isocurvature
contribution is subdominant.

Finally, it has been noted \cite{Linde:1996gt} that this class of
models suffer from the boundary condition of $\langle\hat{X}\rangle=0$
being an unnatural expansion point of the fluctuations of $X$ for
$m_{X}\ll H$. Although it certainly is true that in this limit the $H$
dependent radiative corrections lift the flatness of the potential,
there are no radiative tadpoles that are generated.  Furthermore,
although it is true that once the non-decaying mode decoheres as the
wavelength is stretched outside the horizon, acting like a classical
background with $\langle\hat{X}\rangle\neq0$ over the patch of the
size of that wavelength, there is no strong tuning in choosing a patch
that has $\langle \hat{X}\rangle=0$ as long as the inflation did not
last many orders of magnitude in efolds longer than what is needed to
explain the CMB data. Indeed, because of the slight blue tilt, there
is no infrared divergence in this class of models.  As we will see,
the blue tilt of significant isocurvature amplitudes is still
compatible with observations.

For completeness, let us also explicitly state the cutoffs of our
theory. Since the spectral index of the correlator (\ref{eq:Px}) is
related to $3-3 \sqrt{1-4m_X^4/9H^2}$ and our scenario has
$m_{X}^{2}>0$, the correlator has a blue spectrum (as we will see
explicitly in Sec.~\ref{sec:twopoint}), and thus the loop integrals
(see for example Eqs.~(\ref{eq:twopointfunction}) and
(\ref{eq:approxIsocurvature})) for the two-point
function of isocurvature is independent on IR cut-off. Therefore, an
IR cut-off is not necessary unlike the correlators in the massless
case \cite{Lyth:2007jh}. The UV cutoff of our theory is set to be the
horizon scale at the end of inflation: $k_{\rm UV}=H_{e}a_{e}$.

Finally, we note that we use the scalar metric perturbation
parameterization\begin{equation}
ds^{2}=(1+E)dt^{2}-2a\partial_{i}Fdtdx^{i}-a^{2}\left[\delta_{ij}+A\delta_{ij}+\partial_{i}\partial_{j}B\right]dx^{i}dx^{j}.\label{eq:scalarmetricperturbationconvention}\end{equation}
We make the usual choice for the gauge invariant variable that
describes the inflaton dynamical degree of freedom: \begin{equation}
  \zeta\equiv\frac{A}{2}-H\frac{\delta\rho_{\phi}}{\dot{\bar{\rho}}_{\phi}}.\end{equation}
In terms of this variable, the nearly scale invariant slow-roll
inflaton power spectrum is given as\begin{equation}
\Delta_{\zeta}^{2}(k)\equiv\frac{k^{3}}{2\pi^{2}}P_{\zeta}(k),\end{equation}
where to leading order in the slow-roll parameter \begin{equation}
  \epsilon(\phi_k)=\frac{M_{p}^{2}}{2}\left(\frac{U'(\phi_{k})}{U(\phi_{k})}\right)^{2},\end{equation}
we have \begin{equation}
  P_{\zeta}(k)=\frac{1}{12k^{3}\epsilon(\phi_k)}\frac{U(\phi_{k})}{M_{p}^{4}}.
\end{equation}
In the above, $\phi_{k}$ denotes the field value when the mode $k$ leaves
the horizon. We will give more details about the $X$ fluid variable
when discussing the two-point function in the next section.

In summary, the class of dark matter models that is relevant for this
paper corresponds to cases with gravitationally produced bosonic dark
matter that never fully thermalizes with the reheating radiation
produced from the inflaton decay. The slow-roll inflationary model
produces the dominant curvature perturbation spectrum and couples to
the dark matter sector only gravitationally.

\section{Two-Point Function}
\label{sec:twopoint}
\newcommand{\rhoX}{\rho_{0}^{(p)}}

Although isocurvature perturbations have been previously computed for this
class of models \cite{Chung:2004nh}, here we redo the analysis with
more careful attention to cross-correlations between the curvature
perturbations and the isocurvature perturbations because the
observational constraints have become increasingly stringent.

To begin, consider the energy momentum tensor of $X$:
\begin{equation}
T_{\mu\nu}^{(X)}=\partial_{\mu}X\partial_{\nu}X-g_{\mu\nu}\left[\frac{1}{2}\partial_{\alpha}X\partial^{\alpha}X-V(X)\right]\end{equation}
 where $V(X)=m_{X}^{2}X^{2}/2$. Comparing this to\begin{equation}
T_{\mu\nu}^{\rm (perfect\; fluid)}\equiv(\rho_{X}^{(p)}+P_{X}^{(p)})u_{\mu}u_{\nu}-g_{\mu\nu}P_{X}^{(p)},\end{equation}
we see that if we define \cite{Tabensky} \begin{equation}
u^{\mu}\equiv\frac{\partial^{\mu}X}{\sqrt{g^{\alpha\beta}\partial_{\alpha}X\partial_{\beta}X}},\label{eq:fluidvelocity}\end{equation}
we can satisfy the equality\begin{equation}
T_{\mu\nu}^{(X)}=T_{\mu\nu}^{\rm (perfect\; fluid)},\end{equation}
 if
\begin{eqnarray}\label{eq:rhoXorig}
\rho_{X}^{(p)} \equiv u^{\mu}u^{\nu}T_{\mu\nu}^{\rm (perfect\; fluid)}
=
\frac{1}{2}\partial^{\alpha}X\partial_{\alpha}X+V(X)\label{eq:rhoxenergydensity}\end{eqnarray}
and\begin{equation}
P_{X}^{(p)}\equiv\frac{1}{2}\partial^{\alpha}X\partial_{\alpha}X-V(X).\label{eq:Pressure}\end{equation}
We note that to identify Eq.~(\ref{eq:fluidvelocity}) with the fluid
velocity, $\partial_{\mu}X$ has to be timelike. This is consistent
with the fact that any wave packet made of on-shell 1-particle states
can be decomposed in terms of mode functions characterized by timelike
4-momenta. Unlike the coordinate dependent $T_{00}^{(X)}$,
$\rho_{X}^{(p)}$ is a diffeomorphism scalar. We also note that even
though Eq.~(\ref{eq:rhoxenergydensity}) looks like it has the wrong
sign between the $(\partial_{0}X)^{2}$ and $|\vec{\nabla}X|^{2}$, the
sign is correct and $\rho_{X}^{(p)}$ is positive definite whenever the
fluid interpretation is valid (whenever $\partial_{\mu}X$ is timelike).

We now quantize $X$ by promoting it to an operator $\hat{X}$.
As explained in
  Appendix~\ref{sec:justificationofthebackground}, this allows us to
  identify\begin{equation}
    \label{eq:normalordering}
  \rho_{0}^{(p)}=\langle:\hat{\rho}_{X}^{(p)}:\rangle, \label{eq:steadystatedensity}\end{equation}
  where the normal ordering is with respect to the operators defining
  the $\hat{X}$ vacuum state during the quasi-dS era. After Bogoliubov
  transforming $:\hat{\rho}_{X}^{(p)}:$ to operators of 1-particle
  states relevant for non-dS spacetime,
  $\langle:\hat{\rho}_{X}^{(p)}:\rangle$ will develop a nonzero value
  at that later time. We note that $\langle:\hat{\rho}_{X}^{(p)}:\rangle$
  is homogeneous as long as the vacuum state is spatially translation
  invariant. (Here the inflaton/scalar perturbations are treated as operators, which means that as long as the vacuum governing these are
  spatially translation invariant,
  $\langle:\hat{\rho}_{X}^{(p)}:\rangle$ will be spatially translation
  invariant as well.)
Next, we consider the semi-classical variable\begin{equation}
\delta\rho_{X}^{(p)}\equiv\rho_{X}^{(p)}-\rho_{0}^{(p)}(t).\end{equation}
We can then define the usual fluid variable associated with $\rho_{X}$\begin{equation}
\zeta_{X}\equiv\frac{A}{2}-H\frac{\delta\rho_{X}^{(p)}}{\frac{d}{dt}\rho_{0}^{(p)}(t)},\end{equation}
where we parameterize the spatial scalar metric perturbation as
be
$h_{ij}^{S}=-a^{2}(t)(A\delta_{ij}+\partial_{i}\partial_{j}B)$ with $\bar{g}_{\mu\nu}=\mbox{diag}\{1,-a^{2},-a^{2},-a^{2}\}$.
Under the diffeomorphism $t\rightarrow t-\xi^{0}$, we have
\begin{eqnarray}
A & \rightarrow & A+2H\xi^{0}\\
\delta\rho_{X}^{(p)}&\rightarrow &\delta\rho_{X}^{(p)}+\xi^{0}\frac{d}{dt}\rho_{0}^{(p)}\label{eq:deltarhoXtransform},\end{eqnarray}
which makes $\zeta_{X}$ first order gauge invariant, as expected. Similarly, we
can define gauge invariant variables $\zeta$, $\zeta_{\phi}$, and $\zeta_{R}$
associated with  total energy density $\rho$, inflaton energy density
$\rho_{\phi}$, and radiation energy density $\rho_{R}$, respectively. Since the
gauge invariant isocurvature variable that describes the difference between the
dark matter and the radiation, which is an inflaton descendant, is
conventionally defined as \cite{Wands:2000dp}
\begin{equation}
\delta_{S_{X}}\equiv3\left(\zeta_{X}-\zeta_{R}\right).
\end{equation}
Note that the inflaton eventually decays into radiations and
matters, while curvature perturbation $\zeta \approx \zeta_{\phi}$ remains a conserved
quantity on long wavelengths even after the inflaton decay and it is
adiabatically matched to $\zeta_{R} \approx \zeta$, since the dark matter is
energetically subdominant at the primordial epoch.

In the comoving gauge defined by the coordinate system in which the
inflaton fluctuations vanish (i.e., $\delta\phi=0$), we have
$\zeta_{R}=\zeta_{\phi} =A^{(c)}/2$. Hence,
\begin{equation}
\delta_{S_{X}}=\delta_{X}^{(c)}=\frac{\delta\rho_{X}^{(p)(c)}}{\rho_{0}^{(p)}(t)},\end{equation}
where the $(c)$ superscript refers to the comoving gauge quantity and
we have used the fact that $\rho_{0}$ behaves as $a^{-3}$ once the
Hubble scale is sufficiently smaller compared to the mass $m_{X} \ll
H$. Therefore, correlator combinations involving $\delta_{X}^{(c)}$
and $A^{(c)}$ are of primary physical interest. To compute them, we
quantize $\delta\rho_{X}^{(p)(c)}$ by promoting
$\rho_{X}^{(p)(c)}\rightarrow\hat{\rho}_{X}^{(p)(c)}$ (through the
quantization of $\hat{X}$) and promoting $\rho_{0}^{(p)}(t)$ to an
identity operator (since this was already defined semi-classically as
a matrix element according to Eq.~(\ref{eq:background})) as
follows:\begin{equation}
  \delta\hat{\rho}_{X}^{(p)(c)}=:\hat{\rho}_{X}^{(p)(c)}:-\hat{1}\rho_{0}^{(p)}(t).\end{equation}
This can be used to compute
$\langle\delta\hat{\rho}_{X}^{(p)(c)}(t,\vec{r})\delta\hat{\rho}_{X}^{(p)(c)}(t,0)\rangle$
and then can be divided by $\rho_{0}^{2}(t)$ after this quantity
settles down to give an expression for
$\langle\hat{\delta}_{X}^{(c)}(t,\vec{r})\hat{\delta}_{X}^{(c)}(t,0)\rangle$.

\begin{figure}
  \begin{centering}
   \includegraphics[scale=0.5]{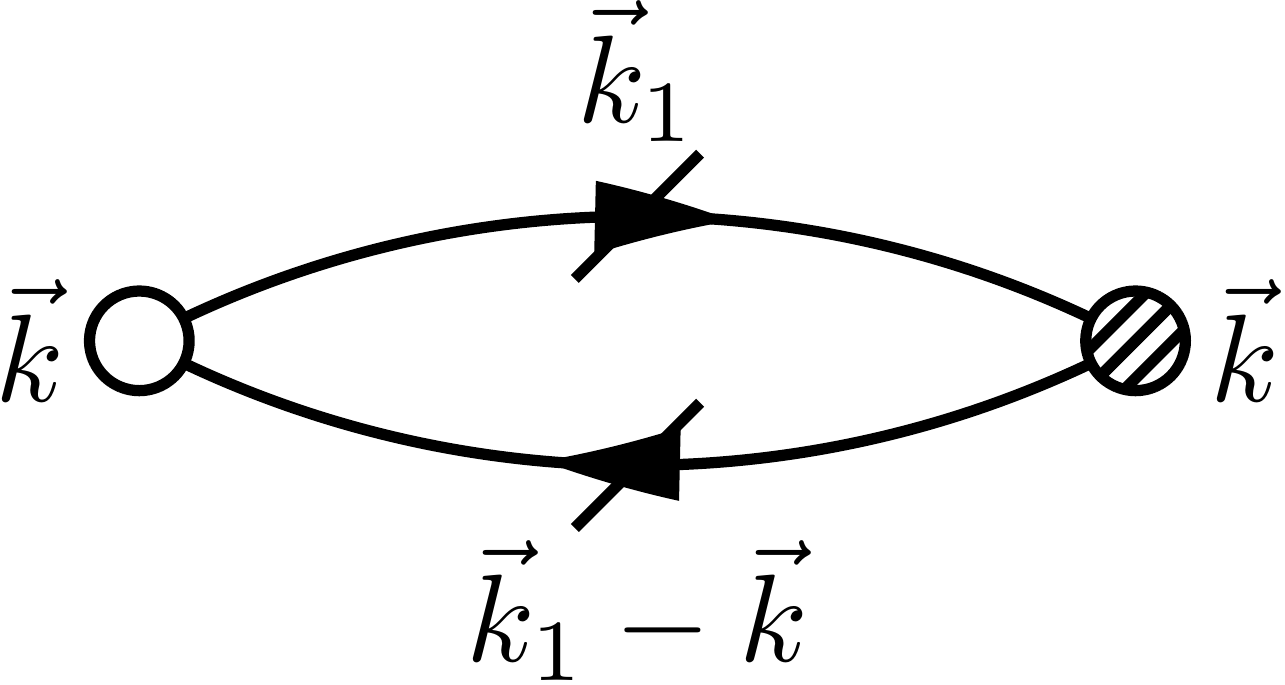}
\par\end{centering}

\caption{\label{fig:2-point}A diagrammatic representation of the
  2-point function of a composite operator $\delta_{X}$.  The open
  circle corresponds to external momentum flowing out, and the shaded
  circle to external momentum flowing in. The slashes on the legs
  indicate that the $X$ propagators are on-shell.}

\end{figure}

Diagramatically, the 2-point function is as shown in Fig.~\ref{fig:2-point}.
Hence, the 2-point power spectrum is\begin{eqnarray}
\label{eq:twopointfunction}
\Delta_{\delta_{S_{X}}}^{2}(k) & = & \frac{k^{3}}{2\pi^{2}}\int d^{3}re^{-i\vec{k}\cdot\vec{r}}\langle\hat{\delta}_{X}^{(c)}(t,\vec{r})\hat{\delta}_{X}^{(c)}(t,0)\rangle \nonumber \\
 & \approx & \frac{k^{3}}{2\pi^{2}}2\int^{a_{e}H_{e}}_{\Lambda_{IR}}\frac{d^{3}k_{1}}{(2\pi)^{3}}P_{X}(k_{1})P_{X}(|\vec{k}-\vec{k}_{1}|),\end{eqnarray}
in which
\begin{equation}
\label{eq:Px}
P_{X}(k)\equiv\frac{m_{X}^{2}}{2\rhoX}|X_{k}|^{2},\end{equation}
$\hat{\delta}_{X}^{(c)}$ is approximated as $m_X^2\hat{X}^2/2\rhoX$,
and $X_{k}$ is the solution to the mode equation in
Appendix~\ref{sec:modefunc}.
For $m_{X}/H<3/2$ and $k/(aH)\ll1$, we
can express $P_{X}(k)$ approximately as \begin{equation}
\label{eq:PxApprox}
  P_{X}(k)=A_{X}\left(\frac{k}{k_{0}}\right)^{\gamma_{X}(H_{*})}k^{-3},\end{equation}
where \begin{equation}
\gamma_{X}(H)=3-3\sqrt{1-\frac{4m_{X}^{2}}{9H^{2}}}>0.\end{equation}
In \cite{Chung:2004nh}, $H_{*}$ is allowed a wave
vector dependence; however, this effect is a subdominant correction
to the already small $(m_{X}/H_{*})^{2}$. It is important to note
that for the parameter range of interest $\gamma_{X}\ll1$ such that
$k^{3}P_{X}$ is nearly scale invariant, the amplitude $A_{X}$ is
given by the approximate formula \begin{equation}
A_{X}\sim\frac{10^{2}|\Gamma(\frac{3}{2}-\frac{\gamma_{X}(H_{*})}{2})|^{2}}{2^{\gamma_{X}(H_{*})}\pi}\frac{m_{X}H_{*}^{2}}{H_{e}^{3}}\exp\left[\frac{1}{\epsilon}\Re\left\{ \gamma_{X}(H_{*})-\gamma_{X}(H_{e})+3\ln\left(\frac{1-\frac{1}{3}\gamma_{X}(H_{*})+\frac{H_{*}}{m_{X}}}{1-\frac{1}{3}\gamma_{X}(H_{e})+\frac{H_{e}}{m_{X}}}\right)\right\} \right],\end{equation}
where $\epsilon=\frac{M_{p}^{2}}{2}\left(\frac{U'(\phi)}{U(\phi)}\right)^{2}$
is the usual inflaton slow-roll parameter. The complicated exponential
factor arises from considering the time evolution of the mode function
$X_{k}$ to a time beyond the time when $k/a<\epsilon/H_{*}$. We see that with
a tuning of the mass parameter to $O(0.1)$ precision, $A_{X}$ can
be a small number despite the exponential factor containing $\epsilon^{-1} \gg 1$.
For the dark matter abundance to be compatible with cosmological observations,
it is important that $H_{*}>H_{e}$ while $m_{X}/H_{e}\sim O$(1).
Using these approximations and Eq.~(\ref{eq:PxApprox}), we find that
Eq.~(\ref{eq:twopointfunction}) yields \begin{eqnarray}\label{eq:approxIsocurvature}
\Delta_{\delta_{S_{X}}}^{2}(k) & \approx &
\frac{k^{3}}{2\pi^{2}}A_{X}^{2}\left(\frac{k}{k_{0}}\right)^{2\gamma_{X}}\frac{1}{k^{3}}\times
2\int^{a_{e}H_{e}/k}_{\Lambda_{IR}/k}\frac{d^{3}u}{(2\pi)^{3}}u^{\gamma_{X}-3}\left|1-\vec{u}\right|^{\gamma_{X}-3}
\nonumber \\
& \approx &
\frac{4}{\gamma_{X}}\left[1-\left(\frac{\Lambda_{IR}}{k}\right)^{\gamma_{X}}\right]\left[\frac{A_{X}}{2\pi^{2}} \left(\frac{k}{k_{0}}\right)^{\gamma_{X}}\right]^{2}\nonumber \\
& \approx &\frac{4}{\gamma_{X}}\left(\frac{k^{3}}{2\pi^{2}}P_{X}(k)\right)^{2},\label{eq:deltaSpowerspec}\end{eqnarray}
where we have used that $\gamma_{X} \ll 1$ in the second line. Note that IR cutoff $\Lambda_{IR}$ dependence does not appear because
of the blueness of $P_{X}$ (i.e. $\gamma_{X}(H_{*})>0$), and $\Delta_{\delta_{S_X}}^2$
inherit a spectrum of $k^{3}P_{X}(k)$. On the other hand, we will see
in the next section that the key to obtaining large non-Gaussianities is
that $\left(k^{3}P_{X}(k)\right)^{2}$ is much smaller than $k^{3}P_{X}(k)$.

Thus far, we have focused on only the $X$ isocurvature
perturbations. In the mixed scenario described near
Eq.~(\ref{eq:mixed}), we can rescale Eq.~(\ref{eq:deltaSpowerspec}) to
obtain the total CDM isocurvature perturbations as
\begin{equation}\label{eq:relationdSanddSx}
  \Delta_{\delta_{S}}^{2}(k)=\omega_{X}^{2}\Delta_{\delta_{S_{X}}}^{2}(k),\end{equation}
since the rest of the dark matter contribution has no isocurvature
perturbations.

\begin{figure}
  \begin{centering}
   \includegraphics[scale=0.3]{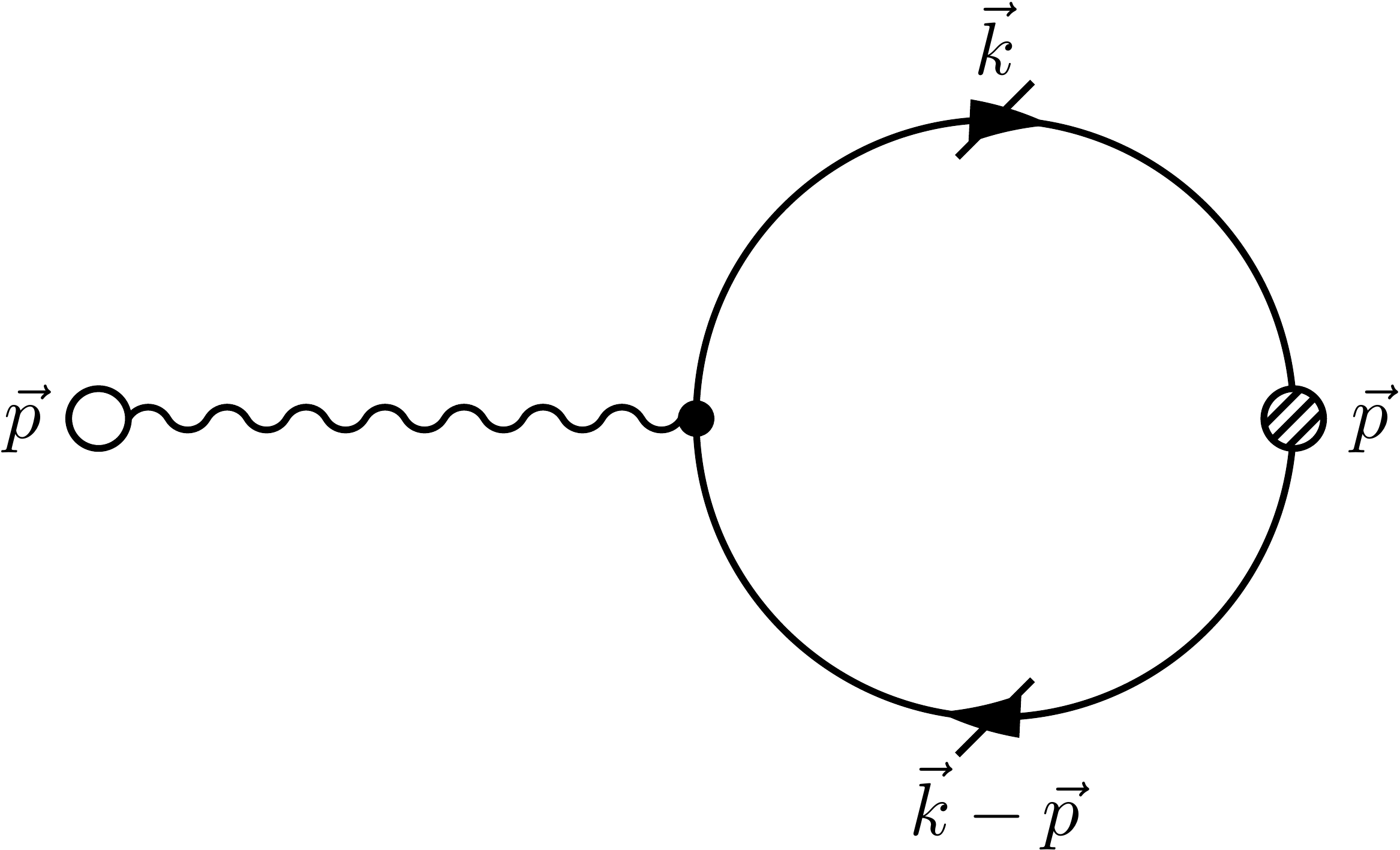}
\par\end{centering}

\caption{\label{fig:crosscorr}A diagrammatic representation of the
  cross-correlator $\langle \zeta \delta_S\rangle$, where the leading
  interaction vertex is proportional to $m_X^2$.  The wavy propagator
  corresponds to the on-shell $\langle \zeta \zeta \rangle$
  correlator.  The open circle corresponds to external momentum flowing
  out, and the shaded circle to external momentum flowing in. The
  slashes on the legs indicate that the $X$ propagators are on-shell.}

\end{figure}

%


As we will see in the next section, the bispectrum is maximized in the
parameter region in which $\Delta_{\delta_{S_{X}}}^2(k)$ is large.
The observational bound on $\Delta_{\delta_{S}}^2$ is stringent
unless the cross-correlation between curvature and isocurvature is negligible, i.e.
\cite{Komatsu:2008hk,Hikage:2008sk,Sollom:2009vd,Valiviita:2009bp} \begin{equation}\label{eq:powSzeta}
  \Delta_{\zeta \delta_{S}}^{2} \ll \Delta_{\zeta} \Delta_{\delta_{S}},
\end{equation} where $\Delta_{\zeta \delta_{S}}^{2}$ is the power spectrum of
the cross-correlation.
The left hand side $\langle\hat{\zeta}\hat{\delta}_{S}\rangle$ corresponding to the diagram shown in
Fig.~\ref{fig:crosscorr} can be computed using the in-in formalism
\cite{Maldacena:2002vr,Weinberg:2005vy} using the trilinear
interaction Hamiltonian in the comoving gauge
\begin{equation}
H_I(t) \ni - \int d^3x\, a^{3}(t)\left[
 \hat{T}^{ij}_{X}(t,\vec{x}) a^{2}(t) \delta_{ij} \hat\zeta(t,\vec{x})\right],
\end{equation}where $T^{\mu\nu}_{X}$ is the stress energy tensor of $X$. Note that
other interaction Hamiltonian contributions are derivatively
suppressed. As will be shown elsewhere \cite{crosscorr-inprep}, the
curvature-isocurvature cross correlation is
\begin{equation}
\beta\equiv\frac{\Delta_{\zeta \delta_{S}}^{2}}{\Delta_{\zeta} \Delta_{\delta_{S}}}
\lesssim \frac{\Delta_{\zeta}}{2} \sim 2.5\times 10^{-5},
\end{equation}which shows that the cross-correlation is negligible.
This fact is understood by the soft-$\zeta$
theorem\cite{Maldacena:2002vr,Hinterbichler:2012nm}, which allows to factorize $\left\langle \hat{\zeta}\hat{\zeta}\right\rangle$
from $\left\langle \hat{\delta\rho_{X}}\hat{\zeta}\right\rangle$, i.e.
\begin{equation}
\left\langle \delta \hat{\rho}_{X}\hat{\zeta}_{\vec{p}}\right\rangle \sim
\left\langle \hat{\zeta}_{-\vec{p}} \hat{\zeta}_{\vec{p}} \right\rangle
\frac{\partial}{\partial \ln a} \left\langle \hat{\rho}_{X}\right\rangle
\end{equation} up to a momentum conserving delta function.  Physically, the curvature perturbation $\zeta$
can affect the energy density $\rho_{X}$ and generate correlation only
at its horizon crossing, because after the perturbation $\zeta$
crosses the horizon and then freezes, it can be effectively treated as
a gauge mode, which corresponds to the spatial dilation in the general
coordinate transform.

Because the isocurvature is of the
uncorrelated type in this scenario, the
current observational bound on the adiabaticity in terms of the
parameter $\alpha$ from WMAP, BAO, and SN combined \cite{Komatsu:2008hk,Komatsu:2010fb} becomes
\begin{equation}
\alpha =\frac{\Delta_{\delta_S}(k_{0})}{\Delta_\zeta (k_{0})+\Delta_{\delta_S}(k_{0})}< 0.064,
\end{equation}
where $k_{0}=0.002 \mbox{Mpc}^{-1}$.  Note the contraint is considered
under the assumption that the isocurvature perturbation is
scale-invariant, while our model predicts the blue-tiled
spectrum. However, we expect that the bound should not be either
altered significantly (or more severely constrained) since the
spectral index is less than 1.2 within the parameter range of
interest. Furthermore, recent analyses
\cite{Sollom:2009vd,Valiviita:2009bp} show that the best CMB fit
favors a blue-tilted isocurvature perturbation, which relaxes the
isocurvature constraints.  Therefore, we will use the contraint as a
conservative bound.  In the next section, we will
see that this implies the maximum $f_{NL} \approx O(35)$.

\section{Bispectrum}
\label{sec:bispectrum}
\begin{figure}
\begin{centering}
\includegraphics[scale=0.5]{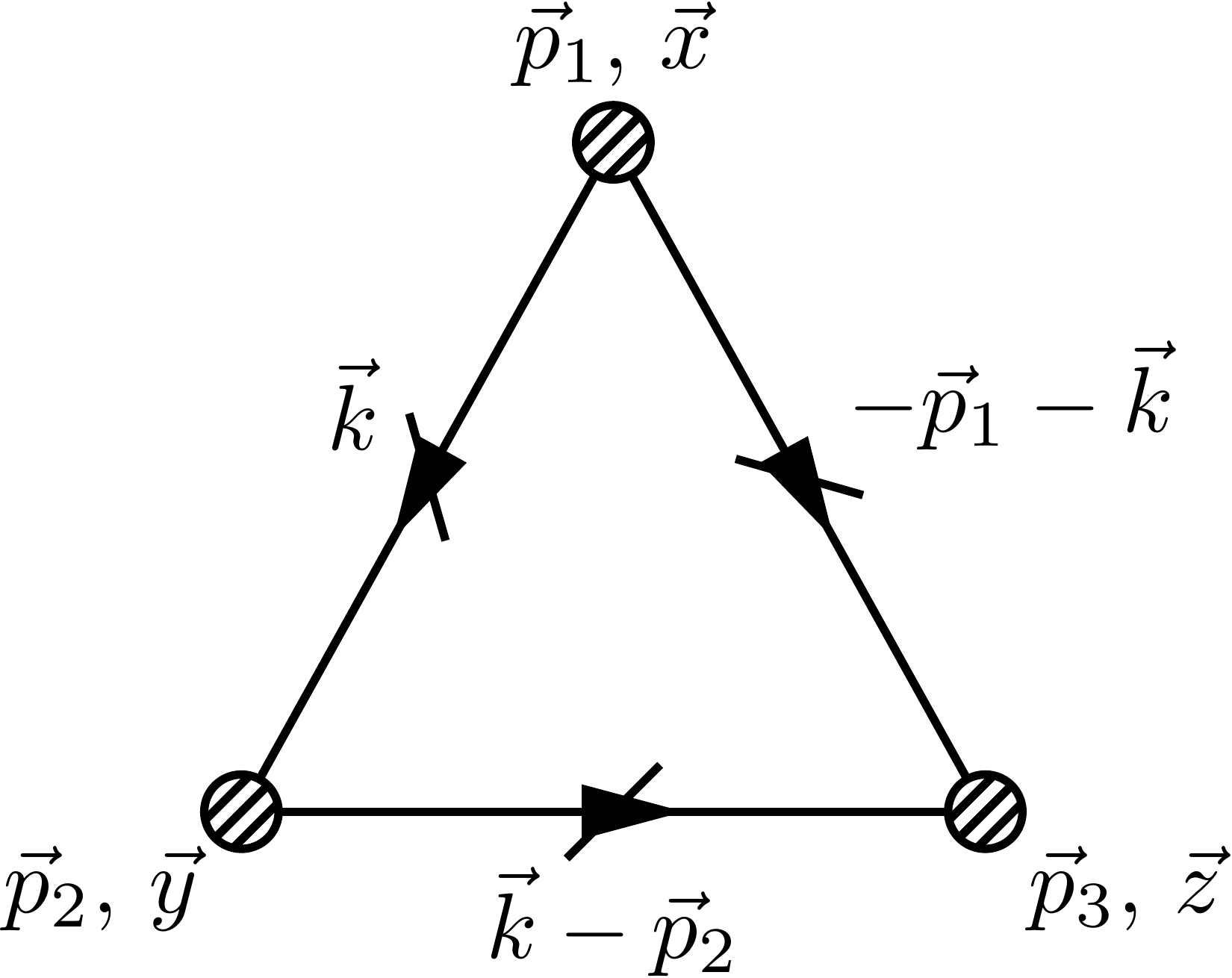}
\end{centering}

\caption{\label{fig:3-point}A diagrammatic representation of the
  3-point function of a composite operator $\delta_{X}$.  The slashes
  indicate that the $X$ propagators are on-shell. The spatial variables
  $\{\vec{x},\vec{y},\vec{z}\}$ indicate the insertion points of the
  external momenta $\vec{p}_{i}$.}
\end{figure}

In this section, we now compute the bispectrum
$B_{\delta_{S}}(\vec{p}_{1},\vec{p}_{2},\vec{p}_{3})$ defined
by \begin{equation}
  (2\pi)^{3}\delta^{(3)}(\sum_{i}\vec{p}_{i})B_{\delta_{S}}(\vec{p}_{1},\vec{p}_{2},\vec{p}_{3})=\omega_{X}^{3}\int
  d^{3}x_{1}d^{3}x_{2}d^{3}x_{3}e^{-i\sum_{n}\vec{p}_{n}\cdot\vec{x}_{n}}\langle\hat{\delta}_{X}^{(c)}(t,\vec{x}_{1})\hat{\delta}_{X}^{(c)}(t,\vec{x}_{2})\hat{\delta}_{X}^{(c)}(t,\vec{x}_{3})\rangle,\end{equation}
  where we recall from the previous section that $\delta_{X}^{(c)}$
  in the comoving gauge can be identified with the gauge invariant
  quantity $\delta_{S_{X}}$. With the diagrammatic representation
  given by Fig.~\ref{fig:3-point}, we find\begin{equation}
  B_{\delta_{S}}(\vec{p}_{1},\vec{p}_{2},-(\vec{p}_{1}+\vec{p}_{2}))\approx8\omega_{X}^{3}\int\frac{d^{3}k}{(2\pi)^{3}}P_{X}(|\vec{k}|)P_{X}(|\vec{p}_{1}+\vec{k}|)P_{X}(|\vec{p}_{2}-\vec{k}|),\end{equation}
  which gives the analytic estimate of the primordial bispectrum for
  different triangle configurations fixed by $\vec{p}_{1}$ and
  $\vec{p}_{2}$.

As large non-Gaussianities are difficult to obtain in slow-roll
inflation in the squeezed triangle limit, we will focus on that limit
in this work. Using the one-pole approximation, we estimate the
isocurvature bispectrum (written in a symmetrized form in the wave
vectors) in the limit that one of the $|\vec{p}_{i}|$ is much smaller
than the others as follows:\begin{equation}
  B_{\delta_{S}}(\vec{p}_{1},\vec{p}_{2},\vec{p}_{3})\approx\frac{8}{\gamma_{X}(H_{*})}\frac{p_{\rm
      min}^{3}}{2\pi^{2}}P_{X}(p_{\rm
    min})\omega_{X}^{3}\left[P_{X}(p_{1})P_{X}(p_{2})+P_{X}(p_{2})P_{X}(p_{3})+P_{X}(p_{3})P_{X}(p_{1})\right]\end{equation}
where $p_{\rm min}={\rm min}\{|\vec{p}_{i}|\}$.
Since the energy density of $X$ is quadratic in $X$, the density
correlator scales as $P_{X}^{2}$ and not just $P_{X}$, which means
that the coefficient $p_{{\rm min}}^{3}P_{X}(p_{{\rm min}})$ is not
quite the power spectrum. Nonetheless, because of the blueness of
$P_{X}$, $p_{{\rm min}}^{3}P_{X}(p_{{\rm min}})$ can be strongly
suppressed if $\gamma_{X}(H_{*})$ is large, and hence
$m_{X}$ has to be smaller than $H_{*}$ for a non-negligible
bispectrum. On the other hand, if $m_{X}$ is too small compared to
$H_{*}$, we saw in the previous section that the isocurvature
perturbations are larger than what is allowed by current data. Hence,
there is a window for which the non-Gaussianities can be large and the
isocurvature perturbations are consistent with the existing data.

To see why the bispectrum composed of quadratic fields is larger than
the bispectrum composed of linear fields (such as for ordinary
inflatons), consider the following ratio of isocurvature bispectrum to
a fiducial local bispectrum defined with $f_{NL}=1$ \begin{equation}
  f_{NL}^{S}\equiv\frac{B_{S}}{\left.B_{\zeta}\right|_{f_{NL}=1}} =
  \frac{5}{6}\frac{B_{\delta_{S}}(\vec{p}_{1},\vec{p}_{2},\vec{p}_{3})}{P_{\zeta}(p_{1})P_{\zeta}(p_{2})+P_{\zeta}(p_{2})P_{\zeta}(p_{3})+P_{\zeta}(p_{3})P_{\zeta}(p_{1})},\end{equation}
where $P_{\zeta}$ is a two-point function of adiabatic
perturbation. On large scales, the $\delta_{S}$ contribution to the
temperature perturbation $\Delta T/T$ compared to the $\zeta$
contribution is different by factor 2 due to the Sachs-Wolfe effect,
i.e. \begin{equation} \frac{\Delta
    T}{T}=-\frac{1}{5}\zeta-\frac{2}{5}\delta_{S}.\end{equation}
However, on scales smaller than $1/k_{eq}$, 
the transfer function of the isocurvature perturbations during
radiation domination is suppressed by an additional factor $k_{eq}/k$
compared to that of adiabatic perturbations.  One can understand this
intuitively in the Newtonian gauge in terms of how isocurvature
perturbations source the gravitational potential which in turn is
proportional to the temperature perturbations.
For adiabatic initial conditions, the gravitational potential on
superhorizon scales are frozen.
In contrast, isocurvature initial conditions effectively fix the
superhorizon gravitational potential to be zero during the early
radiation domination period when the dark matter energy density is
negligible.  As the fraction of dark matter energy density grows
during the radiation domination period, the dark matter perturbations
carrying the isocurvature information source the gravitational
potential until the dark matter becomes the dominant energy component
or until the modes enter the horizon.  Thus, contrary to the
temperature perturbations sourced by the adiabatic perturbations,
those sourced by the isocurvature perturbations are proportional to
the matter fraction at the horizon entry, which yields the additional
suppression factor of $k_{eq}/k$ in the transfer function.

Because the isocurvature transfer function has different
features from the adiabatic one,
the isocurvature
bispectrum leads to CMB temperature imprints distinct from that of the
adiabatic bispectrum.  A careful treatment of the transfer function
incorporating the effects just discussed leads to an approximate
relationship on large scales \cite{Kawasaki:2008sn} which can be
summarized as\footnote{As mentioned in Ref. \cite{Kawasaki:2008sn},
  the isocurvature bispectrum is enhanced by factor 4 instead of 8 due
  to the destructive contribution of the small scale modes. Although
  the scale-invariant isocurvature power spectrum has been used in
  their numerical analysis, their argument still applies to the
  slightly blue-tilted spectrum because the transfer function effect
  of small scales arises from $b_{NL,l}^{(iso)}\equiv\frac{2}{\pi}\int
  dk\,k^{2}g_{Tl}^{(iso)}(k)j_{l}(kr)$, which is independent of the
  spectra index of isocurvature.}
\begin{equation}f_{NL} \approx 4 f_{NL}^{S}\end{equation}
where $f_{NL}$ approximately coincides with the usual local $f_{NL}$
definition.\footnote{This relationship is obtained by comparison
  between the reduced bispectrums $b^{(adi)}_{l_{1}l_{2}l_{3}}$ and
  $b^{(iso)}_{l_{1}l_{2}l_{3}}$ only at large angular scales
  ($l_{1},l_{2},l_{3}\lesssim 10$). Thus, $f_{NL}^{S}$ should not be
  interpreted as $f_{NL}^{local}$ in
  Refs. \cite{Komatsu:2008hk,Komatsu:2010fb}, which is obtained by the
  full analysis involving large and small scales.\label{footnote:fnlclarify}}  Thus, in the
squeezed triangle limit, $f_{NL}$ is analytically estimated to
be\begin{equation}
f_{NL}=\frac{80}{3}\frac{\omega_{X}^{3}}{\gamma_{X}}\left(\frac{p_{{\rm
      min}}^{3}}{2\pi^{2}}P_{X}(p_{{\rm
    min}})\right)\frac{P_{X}(p_{1})P_{X}(p_{2})+P_{X}(p_{2})P_{X}(p_{3})+P_{X}(p_{3})P_{X}(p_{1})}{P_{\zeta}(p_{1})P_{\zeta}(p_{2})+P_{\zeta}(p_{2})P_{\zeta}(p_{3})+P_{\zeta}(p_{3})P_{\zeta}(p_{1})}.\label{eq:effectivefnlval}\end{equation}
The large numerical factor of $80/3$ can be traced to the factor of
$8\times 8/2$ that arises from the product of the contraction
permutations, the relative weight of the $\zeta$ and $\delta_{S}$
contributions to $\Delta T/T$, and the transfer function. As $p_{{\rm
    min}}^{3}P_{X}(p_{{\rm min}})$ is suppressed partly from the
smallness of the $P_{X}$ amplitude as well as the blue tilt (causing
$p_{{\rm min}}^{3}$ to be a suppressor), a large ratio of
$P_{X}/P_{\zeta}$ is required for an unsuppressed $f_{NL}$. As we will
now argue, a $P_{X}/P_{\zeta}$ will arise from the fact that
$\delta_{S}$ is quadratic in $X$.

To understand how to obtain a large $P_{X}/P_{\zeta}$, we first
note that since $\delta_{X}^{(c)}$ is a quadratic functional
of $X$ (see Eq.~(\ref{eq:deltaSpowerspec})), we have \begin{equation}
P_{X}\approx\pi^{2}k^{-3}\sqrt{\gamma_{X}\Delta_{\delta_{S}}^{2}}\frac{1}{\omega_{X}}. \end{equation}
If we define \cite{Bean:2006qz}\begin{equation}\label{eq:definitionalpha}
\frac{\alpha}{1-\alpha}\equiv\frac{\Delta_{\delta_{S}}^{2}}{\Delta_{\zeta}^{2}},\end{equation}
we find for $\alpha\ll1$ that\begin{eqnarray}
\frac{P_{X}}{P_{\zeta}}  \sim  \frac{1}{2}\sqrt{\frac{\alpha\gamma_{X}}{\Delta_{\zeta}^{2}}}\frac{1}{\omega_{X}}\sim  10^{4}\sqrt{\alpha\gamma_{X}}\frac{1}{\omega_{X}}.\label{eq:largeness}\end{eqnarray}
Hence, if $\omega_{X}=1$, as long as $\alpha\gg10^{-8}/\gamma_{X}$
, there is a large ratio of $P_{X}/P_{\zeta}$ because
$\Delta_{\zeta}^{2}\ll\sqrt{\Delta_{\zeta}^{2}}$ and $\Delta_{\delta_{S}}^{2}\propto P_{X}^{2}$.
Combining Eq.~(\ref{eq:effectivefnlval}) and Eq.~(\ref{eq:largeness}),
we see\begin{eqnarray}
\label{eq:fnlbound}
f_{NL} & \sim & 6\times
10^{3}\alpha^{3/2}\sqrt{\gamma_{X}}.\end{eqnarray} Hence, if we can
achieve $\alpha\approx 0.07$ \cite{Komatsu:2008hk} and
$\gamma_{X}\sim0.1$ in our model, we can achieve \begin{equation}
  f_{NL}\approx O(35) \label{eq:maxval}\end{equation} on large
scales. Although this $f_{NL}$ cannot be directly interpreted as
$f_{NL}^{local}$ of Ref.~\cite{Komatsu:2008hk,Komatsu:2010fb} (as
discussed in footnote \ref{footnote:fnlclarify}), a similar result
which is obtained through a full numerical analysis in
Ref. \cite{Hikage:2008sk}
\begin{eqnarray}
f_{NL}^{local} & \approx &
30\left(\frac{\alpha}{0.07}\right)^{3/2}\end{eqnarray} gives a
consistency check to our analytic argument leading to
Eq.~(\ref{eq:maxval}). This level of non-Gaussianity is clearly
observable according to the forecasts of Planck and large scale
structure experiments (see e.g.~\cite{Yadav:2007rk,Verde:2009hy}).
Remarkably, this result is independent of $\omega_{X}$. Hence, even
when the dark matter composition of the $X$ particles is small,
non-Gaussianities associated with $X$ may be observable. We note that
hidden in this analytic estimate is the implicit assumption that
$\alpha$ can remain fixed as $\omega_{X}\rightarrow0$.  However,
Eqs.~(\ref{eq:approxIsocurvature}) and (\ref{eq:relationdSanddSx})
show perturbation theory would break down if $\alpha$ needs to be
fixed as $\omega_{X}\rightarrow 0$ limit is literally taken.  For
example,
to have large local non-Gaussianities, the CDM isocurvature should be
$\delta_S = \omega_X \delta_X \sim \alpha^{1/2} \zeta \sim
10^{-5}$. Hence in the case $\omega_X
\lesssim 10^{-5}$, the perturbativity bound of
$\delta_X<1$ is violated.
In the next section, we will compute the relevant quantities more
precisely in the context of a simple $U(\phi)=m_{\phi}^{2}\phi^{2}/2$
inflationary model.

\section{Numerical Results}
\label{sec:numericalresults}
As shown in Sec.~\ref{sec:bispectrum}, a local non-Gaussianity value
of $f_{NL} \sim 30$ is achievable as long as the parameters of the
model result in $\alpha \approx 0.07$ and $\gamma_X \approx 0.1$.
The identification of these parameters requires the
computation of $\left|X_k\right|^2$ associated with the mode function.
Although an analytic approximation exists in
Appendix~\ref{sec:modefunc}, it is still difficult to identify the
parametric dependence because of the fact that the massive field $X$,
unlike the variable $\zeta$, evolves during inflation even when the
mode wavelength is superhorizon in magnitude. Its evolution depends on
the variable $\gamma_X$, whose slow time dependence is difficult to
account for analytically. Hence, to check the phenomenological
viability of this isocurvature model, we now compute the
necessary mode functions numerically within a $U(\phi) =
m_\phi^2 \phi^2/2$ model.

After inflation, the energy density $\rho_X$ (as investigated
analytically and numerically in \cite{Chung:2001cb,Kuzmin:1998kk}) is
estimated to be
\begin{equation}
  \label{eq:rhoX}
  \frac{\rho_X}{\rho_\phi} \approx 10^{-10} \frac{m_X}{m_\phi}
  \left(\frac{m_\phi}{10^{13}\mbox{ GeV}}\right)^2 \exp(-2\pi m_X/m_{\phi}).
\end{equation}
To obtain the $k$ that appears in the mode function $X_k$, we use the
pivot scale $k_0 = 0.002\text{ Mpc}^{-1}$ and the
standard reheating relationships \cite{Liddle:2003as}
\begin{eqnarray}
  k & = & \frac{a_k}{a_0} k_0 \\
  \frac{a_k}{a_0} & = & \frac{a_k}{a_e} \frac{a_e}{a_0} \\
                  & \approx & 2\times 10^{-31} \left(\frac{a_k}{a_e}
  \right)\left(\frac{H(t_e)}{10^{13} \mbox{GeV}
  }\right)^{-2/3}\left(\frac{T_{\rm RH}}{10^9 \mbox{GeV}}\right)^{1/3}.
\end{eqnarray}
The scale factor ratio $a_k / a_e$ is computed directly from the
solution of $\phi$ with the potential $U(\phi)$.
The mode function $X_k$ is then obtained by solving the equation of motion
\begin{equation}
  \label{eq:kgeq}
  \ddot X_k + 3 H X_k + \frac{k^2}{a^2} X_k + m_X^2 X_k = 0
\end{equation}
with the Bunch-Davies initial condition.

\begin{figure}[htb]
  \begin{center}
    \includegraphics[scale=1]{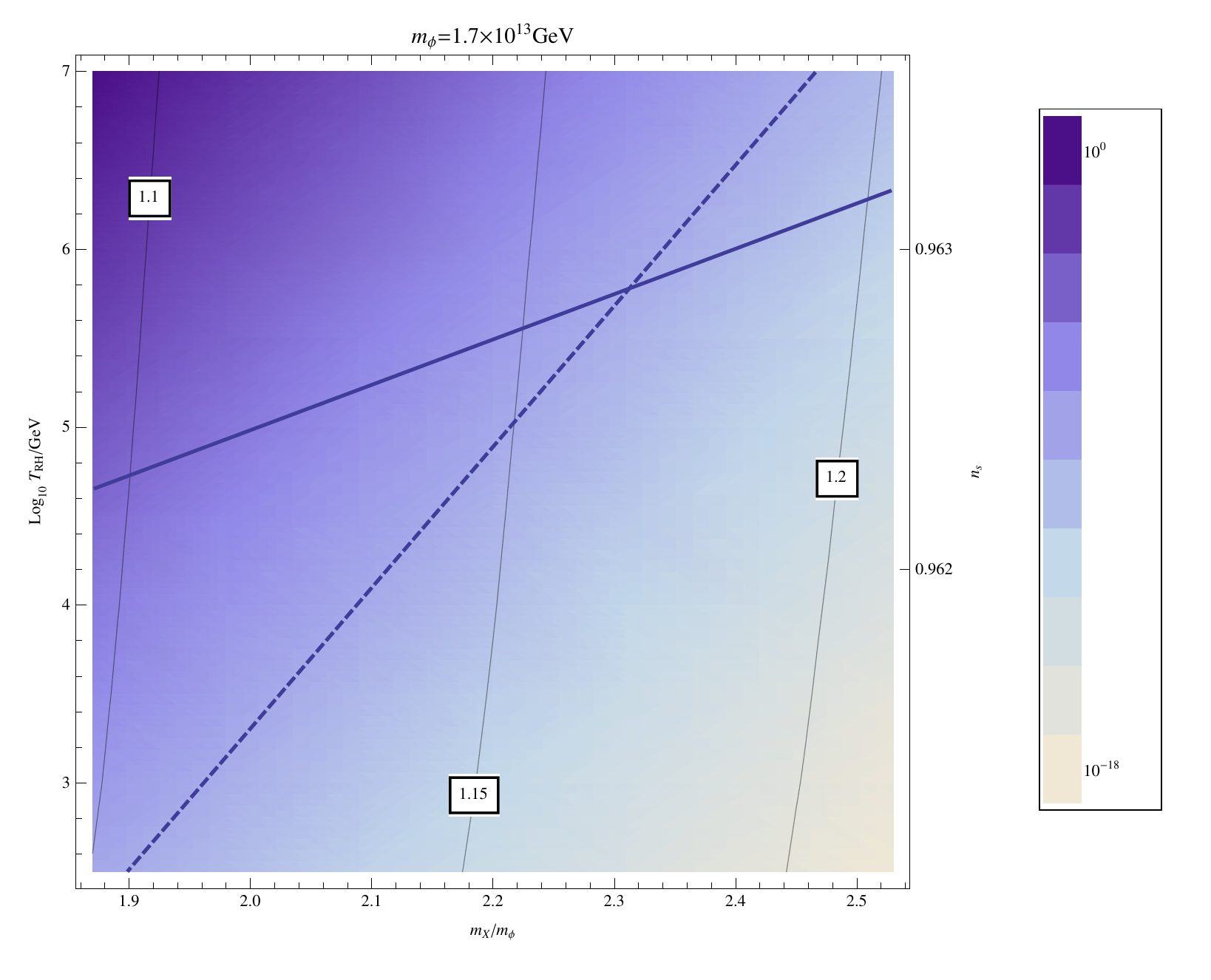}
    \caption{The right bottom of the parameter space under the thick lines
      is allowed by constraints: the current matter density (the solid
      line) and the upper limit ($\alpha \sim 0.07$) of the
      isocurvature power spectrum (the dashed line). The background color
      shows the power spectrum amplitude and the labels on the thin
      lines represents the spectral index of the isocurvature
      $n_X$. The right y-axis shows the spectral index of the
      curvature perturbation $n_s$. }
    \label{fig:parameter_plot}
  \end{center}
\end{figure}

In Fig.~\ref{fig:parameter_plot}, we plot the allowed parameter space
given the isocurvature perturbation and relic abundance constraints. We see that large local non-Gaussianities by the superheavy dark matter
with a small isocurvature power spectrum are attainable in the vicinity
of the thick dashed line of $\alpha = 0.07$. Thus, the upper left
parameter region of the dashed line is ruled out due to overproduction
of isocurvature perturbations. Furthermore, the region above of the
solid line is excluded by the relic abundance condition $\Omega_X <
\Omega_{\text{M}} \lesssim 0.2$.  These conditions for large local
non-Gaussianities yields a robust bound on the reheating temperature
of
\begin{equation}
  T_{\rm RH} \lesssim 10^6\text{ GeV},
\end{equation}
as well as a bound on the mass of the superheavy dark matter of
\begin{equation}
  \frac{m_X}{m_\phi} \lesssim 2.3,
\end{equation}
which corresponds to $m_X \lesssim 4 H_e$. We see once again that
large observable non-Gaussianities can come from dark matter particles
that only compose a small fraction of the total dark matter.  In
Fig.~\ref{fig:parameter_plot}, this region corresponds to the region
along the dashed curve that is far below the solid curve.  This means
that the CMB non-Gaussianities are sensitive to bosonic stable
particles that are negligible as far as their contribution to the
gravitational energy budget today.  As discussed in
Sec.~\ref{sec:bispectrum}, the dashed curve in
Fig.~\ref{fig:parameter_plot} does not continue indefinitely as
$m_X/m_\phi \rightarrow 0$ since perturbation theory for the
primordial spectrum breaks down when $\omega_X$ is less than around
$10^{-5}$.  Parametrically, this breakdown point occurs for
$m_X/m_\phi \approx 1.4$, which corresponds to $m_X \approx 3
H_e$.

From Fig.~\ref{fig:parameter_plot}, we can also see that $\gamma = n_X -1
\approx 0.1$ from the isocurvature spectral index $n_X$. This yields
the advertised result that the isocurvature non-Gaussianities can
reach $f_{NL} \approx 30$.

In this chaotic inflationary model that realizes large
non-Gaussianities, the low reheating temperature $T_{\rm RH} \ll H_e$
implies a long period of matter domination that may lead to nontrivial
dark matter and inflaton condensate clustering phenomenology (see
e.g.~\cite{Carr:1994ar,Berkooz:2005sf,Easther:2010mr}).  As
demonstrated in Sec.~\ref{sec:bispectrum}, a lower inflationary scale
model that still realizes $\alpha \sim O(1)$ and $\gamma \sim O(0.1)$
would allow for an evasion of any phenomenologically undesirable
features that may arise from the long duration of clustering.

\section{Conclusions}
\label{sec:conclusions}
In this paper, we have analyzed the effect of nonthermal dark matter
consisting of weakly interacting gravitationally produced $X$ bosons
on the two and three point functions of the primordial CMB spectrum.
We have demonstrated that large local non-Gaussianities characterized
by $f_{NL} \approx 30$ can result for a particular set of
masses and reheating temperatures without violating isocurvature
bounds.  The conditions that result in large $f_{NL}$ yield a bound on
the reheating temperature of around $10^6$ GeV, as well as a bound on
the dark matter mass of around $3H_e \lesssim m_X\lesssim 4H_e$.  For
lower allowed values of $m_X$ masses, the $X$ bosons can generate a
large $f_{NL}$ value despite the fact that they are an essentially
negligible fraction of the dark matter in the universe.

Although explicit numerical computations were carried out to
find the phenomenologically viable region only in the chaotic
$m_\phi^2 \phi^2/2$ inflationary model, we have presented analytic
arguments to demonstrate that a similar semi-quantitative behavior is
expected for most slow-roll inflationary models, including models with
lower inflationary scales.  The agreement between the analytic
considerations and the numerical results represents a nontrivial
self-consistency check.

The mechanism presented in this paper connects nonthermal dark matter
phenomenology to inflationary phenomenology.  The possibility that a
future discovery of large local non-Gaussianities may provide support
for the existence of a nonthermal dark matter component is indeed
intriguing.  This may even be considered one of the several generic
string phenomenological signatures associated with nonthermal dark
matter components such as those considered in
\cite{Acharya:2008bk,Acharya:2009zt}, although a careful
generalization of our work is required to apply the results in that
context.  We look forward to studying further hints nature may be
willing to reveal in this direction through observations at the
on-going Planck mission and other experiments that probe large scale
structure.

\begin{appendix}

\section{\label{sec:modefunc}Mode functions beyond the dS approximation}

To calculate the correlation function or the energy density, it is
necessary to solve the following mode equation~(\ref{eq:kgeq}).
In dS space, where $H \equiv \dot a / a$ is constant, this
has a well known solution with the Bunch-Davies boundary condition
\begin{eqnarray}
  \label{eq:dssol}
  X_k^{\text{dS}}(t) = \frac{\sqrt{\pi}}{2 a^{3/2} \sqrt{H}} e^{i
    \frac{\pi}{2}(\nu+1/2)} H_\nu^{(1)}\left(\frac{k}{a H}\right),
\end{eqnarray}
in which $\nu =\sqrt{9/4 - m_X^2/ H^2}$ and $H_\nu^{(1)}$ is a Hankel
function.  In the long-wavelength limit, $k/a \ll H$, this solution
behaves as
\begin{eqnarray}
  \label{eq:dssollongwave}
  X_k^{\text{dS}}(t) \approx - \frac{1}{\sqrt{\pi}} (-1)^{3/4}
  2^{-1+\nu} e^{i \nu \pi/2} \left( \frac{k}{a H}\right)^{-\nu}
  \frac{\Gamma(\nu)}{\sqrt{H a^3}},
\end{eqnarray}
in which the decaying solution has been dropped.  For $m_X/H > 3/2$,
$|X_k^{\mbox{ds}}|^2\propto a^{-3}$ dilutes like pressureless dust.

However, in a quasi-dS spacetime in which $H$ varies slowly, this
solution fails to be a good approximation after many efolds past
the horizon crossing: $\Delta t \equiv t-t_* \lesssim 1/ \epsilon H$,
where $\epsilon \equiv -\dot H / H^2$. Conversely, Eq.~(\ref{eq:dssol}) is
a good approximation up to the time $t_*$ just after the horizon
crossing.  (\ref{eq:kgeq}) can be approximated for $t>t_*$ as
\begin{eqnarray}
  \ddot X_k(t) + 3 H(t) X_k(t) + m_X^2 X_k(t) \approx 0,
\end{eqnarray}
and hence we can use the following ansatz for the non-decaying mode
for small $\epsilon \ll 1$.
\begin{eqnarray}
  X_k(t) = X_k^{\text{dS}}(t_*) \left( \frac{a(t_*)}{a(t)} \right)^{3/2}
  \exp\left[ \int_{t_*}^t dt_1\, H(t_1) \nu(t_1) \right].
\end{eqnarray}
Now consider a quasi-dS spacetime with the constant slow-roll
parameter $\epsilon$, $\dot \epsilon = 0$, for which
\begin{eqnarray}
  \frac{1}{H(t)} - \frac{1}{H(t_*)} = \epsilon (t-t_*),\\
  a(t) = a(t_*) (1+ \epsilon H_* t)^{1/\epsilon}.
\end{eqnarray}
Hence, we find
\begin{eqnarray}
  \label{eq:intg}
  \int_{t_*}^{t} dt'\, H(t') \nu(t') & = & \frac{1}{\epsilon} \left[
    \nu(t) -\frac{3}{2} \tanh^{-1}\left(\frac{3}{2 \nu(t)}\right)
    -\nu_* +\frac{3}{2} \tanh^{-1}\left(\frac{3}{2
        \nu_*}\right)\right].
\end{eqnarray}
We note that the mode function $X_k$ is oscillatory for imaginary $\nu$, but that is not reflected in Eq.~(\ref{eq:intg}) because
we have kept only the non-decaying mode in
Eq.~(\ref{eq:dssollongwave}).  To see the oscillatory behavior of the
mode, the decaying mode should be taken into account when the real to
imaginary transition of $\nu$ occurs.  We thus arrive at
\begin{eqnarray}
  | X_k(t) |^2 & \approx & | X_k^{\text{dS}}(t_*) |^2
  \left(\frac{a_*}{a(t)}\right)^3 \exp\left[\frac{2}{\epsilon}
    \Re\left\{\nu(t)-\frac{3}{2}\tanh^{-1}\left(\frac{3}{2\nu(t)}\right)
      -\nu_* +\frac{3}{2} \tanh^{-1}\left(\frac{3}{2 \nu_*}\right)
    \right\}\right] \nonumber\\
  & = & \frac{2^{-2+2\nu_*}}{\pi} \left(\frac{k}{a_*H_*}\right)^{-2
    \nu_*} \frac{\left|\Gamma(\nu_*)\right|^2}{H_* a^3(t)}\exp\left[ \frac{1}{\epsilon} \Re \left\{
      2\nu(t)-2\nu_*\ + 3 \ln \left[
        \frac{\left(\nu_*+\frac{3}{2}\right)
          \frac{H_*}{m_X}}{\left(\nu(t)+\frac{3}{2}\right) \frac{H(t)}{m_X}} \right]
    \right\}\right],
\end{eqnarray}
in which the subscript $*$ indicates the variable is evaluated at $t_*$ and
$\nu_*$ is a positive real number.

\section{Justification of the background}
\label{sec:justificationofthebackground}
In situations in which quantum operators
$\hat{\mathcal{O}}_{\epsilon}^{(q)}$ do not have classical expansion
field values $\mathcal{O}_{0}^{(q)}$, it is necessary to justify how the
$\mathcal{O}_{0}^{(q)}$ are to be identified. In such cases, the typical procedure is to solve for $\hat{\mathcal{O}}_{\epsilon}^{(q)}$ directly in
the presence of linear metric fluctuations because quantum operators
are generically not spatially homogeneous even in Minkowski space
while their matrix elements can be. Hence, it is convenient to
construct $\mathcal{O}_{0}^{(q)}$ from matrix elements of
$\hat{\mathcal{O}}_{\epsilon}^{(q)}$.  Here we outline how to extract
$\mathcal{O}_{0}^{(q)}$ through a spatial average in both the classical
and the quantum case. In the quantum case, the procedure is to construct a
semiclassical quantity that corresponds to the most probable
semi-classical configuration computed from a quantum expectation
value. As explained below, this semiclassical construction is
meaningful when the quantity has a classical interpretation.

Let us define the perturbation $\delta \mathcal{O}^{(q)}$ through
$\mathcal{O}_{\epsilon}^{(q)}= \mathcal{O}_{0}^{(q)}+\delta
\mathcal{O}^{(q)}$.  If $\mathcal{O}_{0}^{(q)}$ is independent of
spatial coordinates for some given fixed coordinate choice and
$\delta\mathcal{O}^{(q)}$ satisfies the condition\begin{equation}
\lim_{L\rightarrow\infty}\frac{1}{V_{L}}\int_{V_{L}}d^{3}x\delta\mathcal{O}^{(q)}=0,\end{equation}
then we have\begin{equation}
\mathcal{O}_{0}^{(q)}\approx\lim_{L\rightarrow\infty}\frac{1}{V_{L}}\int_{V_{L}}d^{3}x\mathcal{O}_{\epsilon}^{(q)}.\label{eq:homogeneous}\end{equation}
Here we do not use a 3-diffeomorphism invariant measure such that
metric perturbations need not be included in the spatial average.
Since unperturbed quantum operators are intrinsically inhomogeneous,
spatial averaging cannot be used to extract $\hat{\mathcal{O}}_{0}^{(q)}$.

On the other hand, for the quantum case, we are only interested in
using it to obtain stochastic boundary conditions.  For a large class
of operators, for any fixed time slice $\Sigma$, there exists a
probability functional $p_{\Sigma}(\{\mathcal{O}_{\epsilon}^{(q)}\})$
such that \begin{equation}
  \langle\hat{\mathcal{O}}_{\epsilon}^{(q_{1})}...\hat{\mathcal{O}}_{\epsilon}^{(q_{r})}\rangle=\int\prod_{q}D\mathcal{O}_{\epsilon}^{(q)}p_{\Sigma}(\{\mathcal{O}_{\epsilon}^{(q)}\})\mathcal{O}_{\epsilon}^{(q_{1})}...\mathcal{O}_{\epsilon}^{(q_{r})},\label{eq:correlatorintermsofprob}\end{equation}
where the left hand side is computed in a fixed vacuum. \footnote{This argument requires generalization when the operators involve time derivatives.  This argument will apply for the case of our interest for correlators of $\hat{X}^2$.} Since we are
computing this at a fixed time, the right hand side functional measure
$D\mathcal{O}_{\epsilon}^{(q)}$ is only discretized over the 3-space
of $\Sigma$. It is important to think of an
$\{\mathcal{O}_{\epsilon}^{(q)}\}$ element in the ensemble (governed
by $p_{\Sigma}$) as a classical configuration only for a fixed time
slice. The time evolution of $\{\mathcal{O}_{\epsilon}^{(q)}\}$ may
not be governed by classical equations starting from that initial
condition. Furthermore, if $p_{\Sigma}$ is sharply peaked, then the
system is essentially in a single configuration (i.e., most elements in
the ensemble are similar). This is one way of reaching classicality.
Suppose there exists a such very probable field configuration on a
fixed time slice. It is convenient to express this probable
configuration (to use for $\mathcal{O}_{\epsilon}^{(q)}$ in
Eq.~(\ref{eq:homogeneous})) using matrix elements.

To construct this probable configuration, we start with the following
question: if there are $N$ samples drawn from a quantum governed
ensemble, how many would have exactly homogeneous configurations
compared to those with inhomogeneous configurations. The set of
exactly homogeneous configurations form a set of measure zero even
though it can be the peak of the $p_{\Sigma}$ functional (unless this
functional diverges at that field configuration point). Most of the
$N$ samples would be inhomogeneous. Having oriented ourselves, the
next question that is relevant for us is what is the most probable
value of Eq.~(\ref{eq:homogeneous}) given $N$ samples in the ensemble
governed by $p_{\Sigma}$. The number of configurations with a given
characteristic $\Gamma$ is given by
\begin{equation}
N_{\Gamma}=N\int\prod_{q}D\mathcal{O}_{\epsilon}^{(q)}p_{\Sigma}(\{\mathcal{O}_{\epsilon}^{(q)}\})\delta(\Gamma)\mbox{det}\frac{\delta\Gamma}{\delta\mathcal{O}_{\epsilon}},
\label{eq:characteristicgamma}\end{equation}
where $\delta(\Gamma)$ represents an appropriately generalized
delta-function in the functional space and the determinant is there
for the appropriate normalization. For a fixed spatial average we have
the regularized constraint\begin{equation}
\Gamma_{\mathcal{O}_{0}^{(q)},V_{L}}\equiv\frac{1}{V_{L}(\Sigma)}\int_{V_{L}(\Sigma)}d^{3}x\mathcal{O}_{\epsilon}^{(q)}-\mathcal{O}_{0}^{(q)},\end{equation}
in which case $N_{\Gamma_{\mathcal{O}_{0}^{(q)},V_{L}}}$ gives the
number of elements in the $N$ sample that realize the homogeneous
value of $\mathcal{O}_{0}^{(q)}$ within the volume $V_{L}$. We will in
the end take $L\rightarrow\infty$ if the infrared regulator removal is
meaningful. Hence, we find\begin{equation}
\frac{\delta\Gamma_{\mathcal{O}_{0}^{(q)},V_{L}}}{\delta\mathcal{O}_{\epsilon}}=\frac{1}{V_{L}(\Sigma)},\end{equation}
which is independent of $\mathcal{O}_{\epsilon}$. The functional
derivatives are taken only with respect to functions of 3-spatial
variable.  Incorporating this into
Eq.~(\ref{eq:characteristicgamma}), we find \begin{equation}
  N_{\Gamma_{\mathcal{O}_{0}^{(q)},V_{L}}}=N\int\prod_{q}D\mathcal{O}_{\epsilon}^{(q)}p_{\Sigma}(\{\mathcal{O}_{\epsilon}^{(q)}\})\delta \left(\frac{1}{V_{L}(\Sigma)}\int_{V_{L}(\Sigma)}d^{3}x\mathcal{O}_{\epsilon}^{(q)}-\mathcal{O}_{0}^{(q)}\right)\frac{1}{V_{L}(\Sigma)}.\end{equation}
Now, the maximum of $N_{\Gamma_{\mathcal{O}_{0}^{(q)},V_{L}}}$ is
obtained by taking a derivative with respect to
$\mathcal{O}_{0}^{(s)}(t)$ (since we are working on a single time
slice, this derivative need not be
functional):\begin{eqnarray} 0 & = &
  \frac{\partial}{\partial\mathcal{O}_{0}^{(s)}(t)}N_{\Gamma_{\mathcal{O}_{0}^{(q)},V_{L}}}\\ &
  = &
  -N\int\prod_{q}D\mathcal{O}_{\epsilon}^{(q)}p_{\Sigma}(\{\mathcal{O}_{\epsilon}^{(q)}\})\delta^{qs}\delta'\left(\frac{1}{V_{L}(\Sigma)}\int_{V_{L}(\Sigma)}d^{3}x\mathcal{O}_{\epsilon}^{(q)}-\mathcal{O}_{0}^{(q)}(t)\right)\frac{1}{V_{L}(\Sigma)},\end{eqnarray}
where the prime on the delta-function corresponds to the derivative
with respect to the functional argument of the delta function. The
functional argument of the delta function can be considered to be a
function of the variation\begin{equation}
\delta\left[\frac{1}{V_{L}(\Sigma)}\int_{V_{L}(\Sigma)}d^{3}x\mathcal{O}_{\epsilon}^{(q)}\right]=\frac{\delta\mathcal{O}_{\epsilon}^{(q)}}{V_{L}(\Sigma)}.\end{equation}
Hence, an integration by parts will yield a solvable equation to the
problem of maximizing $N_{\Gamma_{\mathcal{O}_{0}^{(q)},V_{L}}}$
as\begin{equation}
\int\prod_{q}D\mathcal{O}_{\epsilon}^{(q)}\frac{\delta}{\delta\mathcal{O}_{\epsilon}^{(q)}}p_{\Sigma}(\{\mathcal{O}_{\epsilon}^{(q)}\})\delta\left(\frac{1}{V_{L}(\Sigma)}\int_{V_{L}(\Sigma)}d^{3}x\mathcal{O}_{\epsilon}^{(q)}-\mathcal{O}_{0}^{(q)}\right)=0.\end{equation}
If $\mathcal{O}_{0}^{(q)}$ is chosen to be $\mathcal{O}_{*}^{(q)}$
such that the $\mathcal{O}_{\epsilon}^{(q)}$ configurations that satisfy
\begin{equation}
\mathcal{O}_{*}^{(q)}=\frac{1}{V_{L}(\Sigma)}\int_{V_{L}(\Sigma)}d^{3}x\mathcal{O}_{\epsilon}^{(q)}\label{eq:spatialaverage}\end{equation}
also satisfy \begin{equation}
\frac{\delta}{\delta\mathcal{O}_{\epsilon}^{(q)}}p_{\Sigma}(\{\mathcal{O}_{\epsilon}^{(q)}\})=0,\end{equation}
we have a solution $\mathcal{O}_{*}^{(q)}$. Hence, we must look for
the peak of $p_{\Sigma}(\{\mathcal{O}_{\epsilon}^{(q)}\})$.

Consider\begin{equation}
\langle\hat{\mathcal{O}}_{\epsilon}^{(s)}\rangle=\int\prod_{q}D\mathcal{O}_{\epsilon}^{(q)}p_{\Sigma}(\{\mathcal{O}_{\epsilon}^{(q)}\})\mathcal{O}_{\epsilon}^{(s)}.\label{eq:onepoint}\end{equation}
Assuming $p_{\Sigma}(\{\mathcal{O}_{\epsilon}^{(q)}\})$ is sharply
peaked, consider\begin{equation}
f\equiv\ln p_{\Sigma}(\{\mathcal{O}_{\epsilon}^{(q)}\}).\end{equation}
The peak is located at $\frac{\delta f}{\delta\mathcal{O}_{\epsilon}^{(q)}}=0$
corresponding to the field configuration that satisfies\begin{equation}
\left.\frac{\delta p_{\Sigma}(\{\mathcal{O}_{\epsilon}^{(q)}\})}{\delta\mathcal{O}_{\epsilon}^{(s)}}\right|_{\mathcal{O}_{P}^{(q)}}=0.\end{equation}
Hence, the quadratic expansion of $f$ about the peak configuration
is \begin{equation}
f=f(\mathcal{O}_{P}^{(q)})+\frac{1}{2}\int dx_{1}^{3} dx_{2}^{3} \left.\frac{\delta^{2}f}{\delta\mathcal{O}_{\epsilon}^{(q_{1})}(x_{1})\delta\mathcal{O}_{\epsilon}^{(q_{2})}(x_{2})}\right|_{\mathcal{O}_{P}^{(q)}}\left(\mathcal{O}_{\epsilon}^{(q_{1})}(x_{1})-\mathcal{O}_{P}^{(q_{1})}(x_{1})\right)\left(\mathcal{O}_{\epsilon}^{(q_{2})}(x_{2})-\mathcal{O}_{P}^{(q_{2})}(x_{2})\right),\end{equation}
where the repeated $q_{i}$ indices are summed. $\mathcal{O}_{\epsilon}^{(s)}$ can be raised in Eq.~(\ref{eq:onepoint}) using the usual trick of introducing
a source\begin{equation}
\mathcal{O}_{\epsilon}^{(s)}\rightarrow\frac{\delta}{\delta J}\exp\left[\int
  d^{3}x J\mathcal{O}_{\epsilon}^{(s)}\right]_{J=0}\end{equation}
and carrying out the leading saddle-point approximation integral to obtain\begin{equation}
\langle\hat{\mathcal{O}}_{\epsilon}^{(s)}(x)\rangle=\mathcal{O}_{P}^{(s)}(x),\label{eq:peak}\end{equation}
where we used\begin{eqnarray}
1 & = & \int\prod_{q}D\mathcal{O}_{\epsilon}^{(q)}p_{\Sigma}(\{\mathcal{O}_{\epsilon}^{(q)}\})\label{eq:identity1}\\
 & \approx & e^{f(\mathcal{O}_{*}^{(q)})}\int\prod_{q}D\mathcal{O}_{\epsilon}^{(q)}\times\nonumber \\
 &  & \exp\left\{ \frac{1}{2}\int dx^{3}_{1} dx^{3}_{2} \left.\frac{\delta^{2}f}{\delta\mathcal{O}_{\epsilon}^{(q_{1})}(x_{1})\delta\mathcal{O}_{\epsilon}^{(q_{2})}(x_{2})}\right|_{\mathcal{O}_{P}^{(q)}}\left(\mathcal{O}_{\epsilon}^{(q_{1})}(x_{1})-\mathcal{O}_{P}^{(q_{1})}(x_{1})\right)\left(\mathcal{O}_{\epsilon}^{(q_{2})}(x_{2})-\mathcal{O}_{P}^{(q_{2})}(x_{2})\right)\right\}. \label{eq:identity2}\end{eqnarray}
We note that if we use a spatially translation invariant state to take
the expectation value, $\mathcal{O}_{P}^{(q)}(x)$ is automatically
spatially translation invariant and thus\begin{equation}
\mathcal{O}_{*}^{(q)}=\mathcal{O}_{P}^{(q)}\label{eq:homogeneousisthepeak}\end{equation}
in Eq.~(\ref{eq:spatialaverage}). Hence, Eqs.~(\ref{eq:peak})
and (\ref{eq:homogeneousisthepeak}) combine to give the most probable
spatially averaged configuration to be\begin{equation}
\mathcal{O}_{*}^{(s)}=\langle\hat{\mathcal{O}}_{\epsilon}^{(s)}\rangle\label{eq:background}\end{equation}
to leading order in saddle-point approximation (an expansion in the
peakedness of the distribution function $p_{\Sigma}$).

Hence, when matching to the classical fluid, it is appropriate to
consider the homogeneous background quantity associated with the quantum
operator to be $\langle\hat{\mathcal{O}}_{\epsilon}^{(s)}\rangle$.
It is important to remember that since $\mathcal{O}_{*}^{(s)}$ (in
Eq.~(\ref{eq:spatialaverage})) is being identified with $\mathcal{O}_{0}^{(q)}$
(in Eq.~(\ref{eq:homogeneous})), we are now assuming that the $\epsilon=0$
solution of perturbation theory is governed by equations
that depend only on time for the coordinate system that we have chosen.
\end{appendix}

\section*{Acknowledgments}

We thank Peng Zhou and Toni Riotto for useful conversations.  DJHC
also thanks Xingang Chen and Eichiro Komatsu for discussions following
a preliminary presentation of this work at the {}``Cosmological
Non-Gaussianity Workshop'' which was held on May 13-15, 2011 at the
University of Michigan.  This work was supported in part by the DOE
grant DE-FG02-95ER40896.

\bibliographystyle{apsrev}
\bibliography{references}

\end{document}